# Electrostatics in wind-blown sand

**Jasper. F. Kok[1,2,*] and Nilton. O. Renno[1,2]**

[1]Applied Physics Program, University of Michigan, Ann Arbor, MI, 48109, USA.

[2]Atmospheric, Oceanic, and Space Sciences, University of Michigan, Ann Arbor, MI, 48109.

**ABSTRACT**

Wind-blown sand, or 'saltation,' is an important geological process, and the primary source of atmospheric dust aerosols. Significant discrepancies exist between classical saltation theory and measurements. We show here that these discrepancies can be resolved by the inclusion of sand electrification in a physically based saltation model. Indeed, we find that electric forces enhance the concentration of saltating particles and cause them to travel closer to the surface, in agreement with measurements. Our results thus indicate that sand electrification plays an important role in saltation.

**ARTICLE**

*Introduction.–* In wind-blown sand, or 'saltation,' sand grains are propelled by the wind while they bounce along the surface (Fig. 1). Saltation plays a key role in various geological processes, including wind erosion, sediment transport, and the formation of sand dunes [1,2]. Moreover, the impact of saltating grains on the soil bed is the primary emission mechanism of mineral dust aerosols [2]. These small airborne dust particles affect the Earth system in several ways, including by scattering and absorbing radiation [3], acting as cloud seeds [4], and providing nutrients to ecosystems [5].

Pioneering measurements by Schmidt et al. [6] show that electric fields (E-fields) of up to ~160 kV/m are generated in saltation under moderate wind conditions. Laboratory experiments show that such E-fields facilitate the lifting of sand particles by winds, and can even directly lift sand particles from the surface [7]. Large E-fields are also predicted to occur in Martian saltation, possibly producing large quantities of hydrogen peroxide and making the Martian surface inhospitable to life as we know it [8].

It is known from industrial handling of powders that particle collisions tend to leave smaller particles with net negative charge and larger particles with net positive charge [9]. The physical mechanism governing this charge transfer is poorly understood, although various studies suggest that asymmetric rubbing (i.e., a small area of the small particle rubs over a large area of the large particle) causes a net transfer of electrons from the larger to the smaller particle [10]. We hypothesize that a similar process occurs in saltation, where particles bounce along a soil bed that can be interpreted as the surface of an infinitely large particle (Fig. 1). Saltating particles are thus expected to charge negatively with respect to the surface, as is indeed indicated by measurements of upward-pointing E-fields in wind tunnel and field experiments [6,11]. The occurrence of upward-pointing E-fields in dust devils and dust storms [12] also suggests negatively charged particles over a positively charged surface.

In this Letter, we present the first physically based numerical model of saltation that includes the generation of electric fields and the effects of electric forces on saltation. We show that recent measurements in saltation [13-15] cannot be explained by classical saltation theory [1,16], but are consistent with the predictions of our model when sand electrification is included.

*Theory.–* Saltation is initiated when the wind shear stress $\tau$ exceeds the threshold value $\tau_t$ necessary to move surface particles. The height-integrated particle mass flux $Q$ is a good measure of the intensity of saltation. Experiments show that $Q$ increases approximately cubically with wind shear velocity, $u^* = \sqrt{\tau/\rho_a}$ [1,2], that is

$$Q = Q_0(\rho_a/g)u^{*3}, \qquad (1)$$

where $\rho_a$ is the air density, $g$ is the gravitational acceleration, and $Q_0$ is the dimensionless particle mass flux.

*Model description.–* A detailed description of our model is given in Ref. [17], but we describe it here briefly. We model saltation as the interplay of four processes [18]: (i) the motion of saltating particles, (ii) the modification of the wind profile through momentum transfer with saltating particles, (iii) the collisions of particles with the soil bed, and (iv), the lifting of surface grains by wind stress and by particles impacting the soil bed. The main innovation of our model over previous models (e.g., [18]) is that the charging of saltating particles during collisions with other particles and the surface is included. The effect of the resulting electric forces on particle motion and the threshold shear stress [7] is explicitly accounted for.

We model particle motion in two dimensions by considering gravitational, fluid, and electric forces (Fig. 1). The effects of turbulence and mid-air collisions on particle trajectories are neglected because these effects are relatively small for typical shear velocities [2,18]. Results of laboratory and numerical studies are used to model the collision of particles with the soil bed, including the ejection of surface grains [18].

For the calculation of the wind profile, we make the classical assumption that, in steady-state saltation, the fluid shear stress at the surface stays at the threshold value ($\tau_t$) necessary to initiate the motion of surface particles [16,17]. The particle concentration per unit area is then given by [17]

$$N = \tau_p(0)/\overline{\tau_{sp}(0)} = (\tau - \tau_t)/\overline{\tau_{sp}(0)}, \qquad (2)$$

where $\tau_p(0)$ is the total shear stress exerted by saltating particles at the surface, and $\overline{\tau_{sp}(0)}$ is the average surface stress exerted by a single saltating particle, as computed from the particle trajectories. The size distribution of saltating particles is assumed to be similar to that of the parent soil for particles of 100-500 $\mu$m, in agreement with field measurements [17].

*Sand electrification.-* The model accounts for electrostatic charging of saltating particles during collisions with other particles and the surface. Although collisional charge transfer between grains of granular material is observed in a variety of physical systems [6,11,12], the charging mechanism is not well understood.

Nonetheless, Desch and Cuzzi [19] proposed a model in which the collisional charge transfer depends on the pre-existing charges, the particle sizes, and the difference in the particles' contact potential. They proposed that

$$q_S' = C_1(q_S + q_L) - C_2 \Delta\Phi; \quad q_L' = (1 - C_1)(q_S + q_L) + C_2 \Delta\Phi \qquad (3)$$

where $q_S$ and $q_L$ are the charges of the smaller and larger particles before the collision, $q'_S$ and $q'_L$ are the charges after the collision, $\Delta\Phi$ is the difference in particle contact potential, and $C_1$ and $C_2$ are functions of the mutual capacitances of the two particles, as defined by equations 5-10 of Ref. [19]. For particles of similar composition (i.e., $\Delta\Phi = 0$), such as typical soil particles, (3) suggests that no charge transfer occurs when the colliding particles are not initially charged, which contradicts observations [6,11,12]. To mitigate this problem, we propose an effective contact potential difference between particle pairs of similar composition but different sizes. That is,

$$\Delta\Phi_{eff} = S(r_L - r_S)/(r_L + r_S), \qquad (4)$$

where $S$ (in Volts) is a physical parameter that scales the collisional charge transfer, and $r_S$ and $r_L$ are the radii of the small and large particles. This simple model has a functional form consistent with observations – smaller particles acquire net negative charge during collisions with larger particles, and the charge transfer is reduced as the relative difference in particle size decreases. Since saltating particles impacting the soil surface tend to interact with multiple surface grains [1,18], we interpret the soil bed as the surface of an infinitely large particle (i.e., $r_L = \infty$). By calibrating the model with E-field measurements in saltation [6], we found that $S = 6 \pm 4$ Volts (see Fig. 2).

The soil surface is assumed to be conducting, both because charge exchange with saltating particles provides charge mobility, and because conducting films of water are generally adsorbed on soils [7]. Since the height to which particles saltate is generally much smaller than the horizontal extent over which saltation occurs, we use the infinite plane approximation to determine the electric field $E$ from the calculated space charge density $\rho_c$ and soil surface charge density $\sigma$ [7],

$$E(z) = \frac{1}{2\varepsilon_0}\left(\sigma + \int_0^z \rho_c(z')dz' - \int_z^\infty \rho(z')dz'\right), \tag{5}$$

where $\varepsilon_0$ is the electric permittivity of air. The effect of the surface electric field on the threshold shear velocity ($u*_t$) is calculated using equation 9 in Ref. [7].

*Model results.–* Comparisons of vertical and horizontal profiles of the mass flux in saltation show that predictions of our model are in agreement with measurements (Fig. 3 and Ref. [17]). To the best of our knowledge, the model presented here is the first physically based model capable of accurately reproducing observed particle mass flux profiles. For high shear velocity, the agreement with field measurements improves when electric forces are included (Fig. 3c). The height-integrated mass flux predicted by our model is also in good qualitative agreement with wind tunnel results [17].

Our model predicts that E-fields increase sharply towards the surface and with wind speed (Fig. 2), in agreement with measurements [6,11]. The surface E-field is of particular interest, because E-fields larger than ~80 kV/m reduce significantly the wind shear stress necessary to lift surface particles [7]. Our model finds that this effect contributes to an approximate doubling of the particle concentration due to sand electrification at large shear velocities (Fig. 4).

In addition to increasing the concentration of saltating particles, electric forces also affect particle trajectories. A characteristic height of saltation is the height $z_{50}$ below which 50 % of the mass transport occurs. Classical saltation theory [1,16] predicts that increases in wind speed produce increases in the momentum of saltating particles, causing them to impact and rebound from the surface at higher speed, and therefore reach larger heights. However, recent measurements show that $z_{50}$ stays approximately constant as the wind speed increases [13-15]. This clear discrepancy between measurements and theory can be resolved by the inclusion of sand electrification in our physically based saltation model (Fig. 5). As the negatively charged saltating particles bounce along the positively charged surface (Fig. 1) [6,11], the downward electric force causes particles to travel closer to the surface and at reduced horizontal speed [21]. Since the downward electric force increases with wind speed (Fig. 2), $z_{50}$ remains approximately constant up to moderate shear velocities, in good agreement with measurements [13-15]. At larger shear velocities, electric forces become strong enough to lower the threshold shear velocity [7,17], which reduces the near-surface winds and thus $z_{50}$. We plan to test this prediction with future field measurements.

*Conclusions.–* We developed the first physically based numerical saltation model that includes the effects of sand electrification. Significant discrepancies exist between classical saltation theory and field measurements [13]. We show that the inclusion of sand electrification in saltation models can resolve these discrepancies.

Model results show that sand electrification increases the particle concentration at a given wind shear velocity [7,21] (Fig. 4). Moreover, the downward electric force on saltating particles lowers their trajectories, improving the agreement between model predictions and measurements [13-15] (Fig. 5). Our results thus indicate that sand electrification plays an important role in saltation.

We are also investigating the effect of the classical assumption that the shear stress at the surface remains at the threshold value for particle entrainment [16,17]. Initial results suggest that removing this assumption could also explain some of the discrepancies between theory and measurements [23].

Although this study focuses on terrestrial saltation, preliminary model results suggest that during saltation on Mars, E-fields reach the threshold value of ~25 kV/m for which electric discharges occur [8].

We thank Scott Schmidt for providing his original E-field measurements, and Michael Bretz, William Kuhn, Shanna Shaked, Earle Williams, Daniel Lacks, and an anonymous reviewer for comments. Finally, we thank NSF for financial support through award ATM 0622539.

**REFERENCES**

*Electronic address: jfkok@umich.edu

[1] R. A. Bagnold, *The Physics of Blown Sand and Desert Dunes* (Methuen, London, 1941).

[2] Y. P. Shao, *Physics and Modelling of Wind Erosion* (Kluwer Academic Publishers, Dordrecht, 2000).

[3] I. Tegen, and A.A. Lacis, J. Geophys. Res. **101**, 19237 (1996).

[4] P. J. DeMott et al., Geophys. Res. Lett. **30** (2003).

[5] T. D. Jickells et al., Science 308, **67** (2005).

[6] D. S. Schmidt, R.A. Schmidt, and J. D. Dent, J. Geophys. Res. **103**, 8997 (1998); D. S. Schmidt, personal communication (2007).

[7] J. F. Kok, and N. O. Renno, Geophys. Res. Lett. **33**, L19S10 (2006).

[8] S. K. Atreya et al., Astrobiology **6**, 439 (2006).

[9] I. I. Inculet, G. S. P. Castle, and G. Aartsen, Chem. Eng. Sci. **61**, 2249 (2006).

[10] J. Lowell and W. S. Truscott, J. Phys. D Appl. Phys. **19**, 1281, (1986); D. J. Lacks, and A. Levandovsky, J. Electrostat. **65**, 107 (2007).

[11] X. J. Zheng, N. Huang, and Y. H. Zhou, J. Geophys. Res. **108** (2003); J. J. Qu et al, Sci. China Ser. **D**, 47, 529 (2004).

[12] Freier, G.D., J. Geophys. Res. **65**, 3504 (1960); C. D. Stow, Weather **24**, 134 (1969).

[13] S. L. Namikas, Sedimentology **50**, 303 (2003).

[14] R. Greeley, D. G. Blumberg, and S. H. Williams, Sedimentology **43**, 225 (1996).

[15] K. R. Rasmussen, and M. Sorensen, J. Geophys. Res. (in press).

[16] P. R. Owen, J. Fluid Mech. **20**, 225 (1964).

[17] See EPAPS Document No. … for a detailed model description, as well as model results of the horizontal mass flux profile and the height-integrated mass flux. For more information on EPAPS, see http://www.aip.org/pubservs/epaps.html.

[18] R. S. Anderson, and P. K. Haff, Acta Mech. Suppl. **1**, 21 (1991).

[19] S. J. Desch, and J. N. Cuzzi, Icarus **143**, 87 (2000).

[20] N. Lancaster, J. Sediment. Petrol. **56**, 395 (1986).

[21] The decrease in particle speed due to electric forces reduces the average surface stress exerted by a single saltating particle ($\overline{\tau_{sp}(0)}$ in (2)), which contributes to the increase in particle concentration seen in Fig. 4.

[22] M. P. Almeida, J. S. Andrade, and H. J. Herrmann, Phys. Rev. Lett. **96,** 018001 (2006).

[23] J. F. Kok, and N. O. Renno (to be published).

**FIGURES**

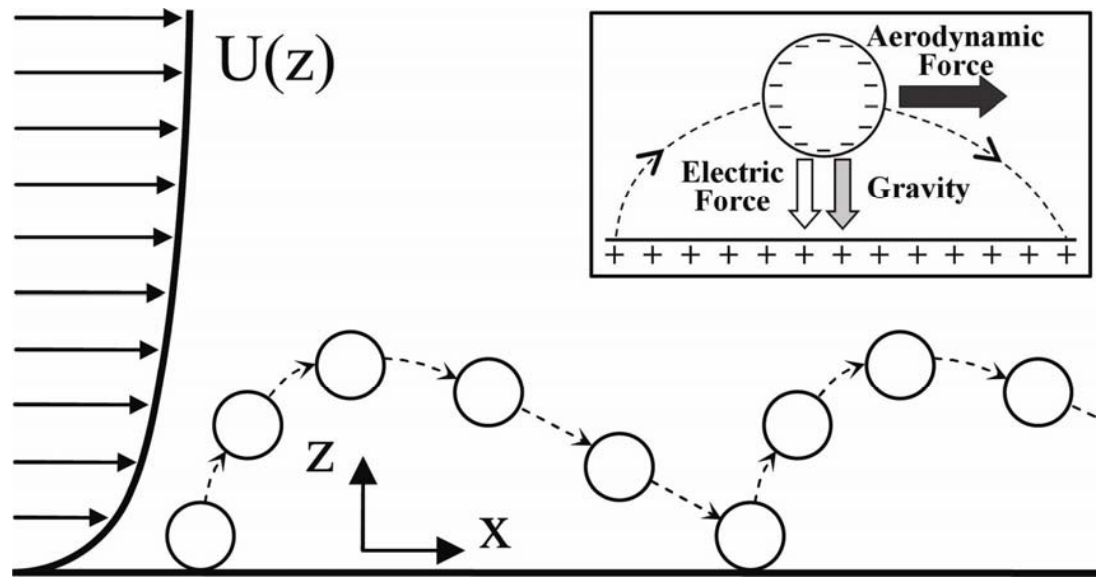

FIG. 1.

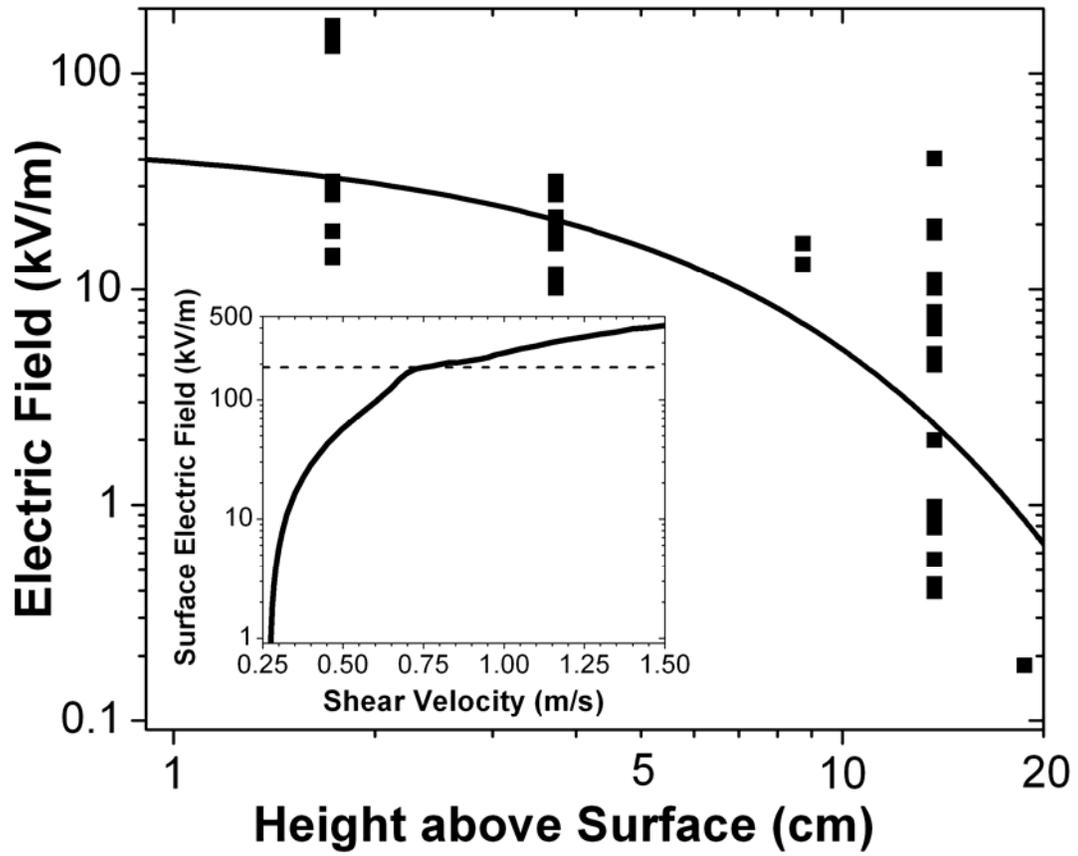

FIG. 2.

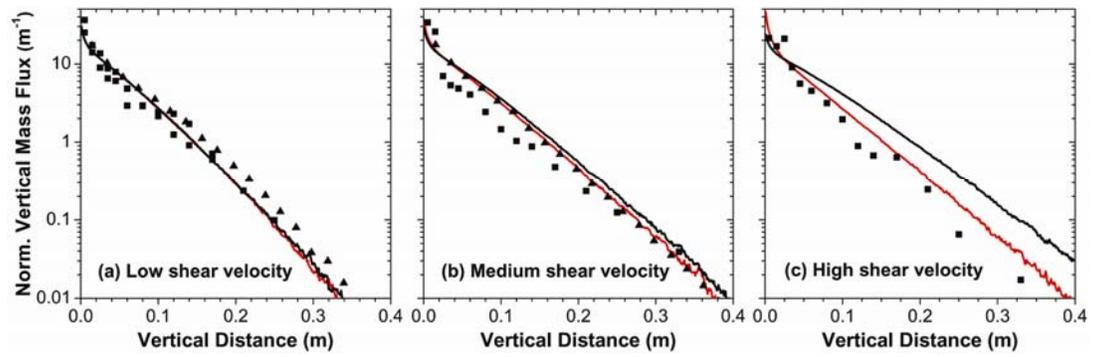

FIG. 3.

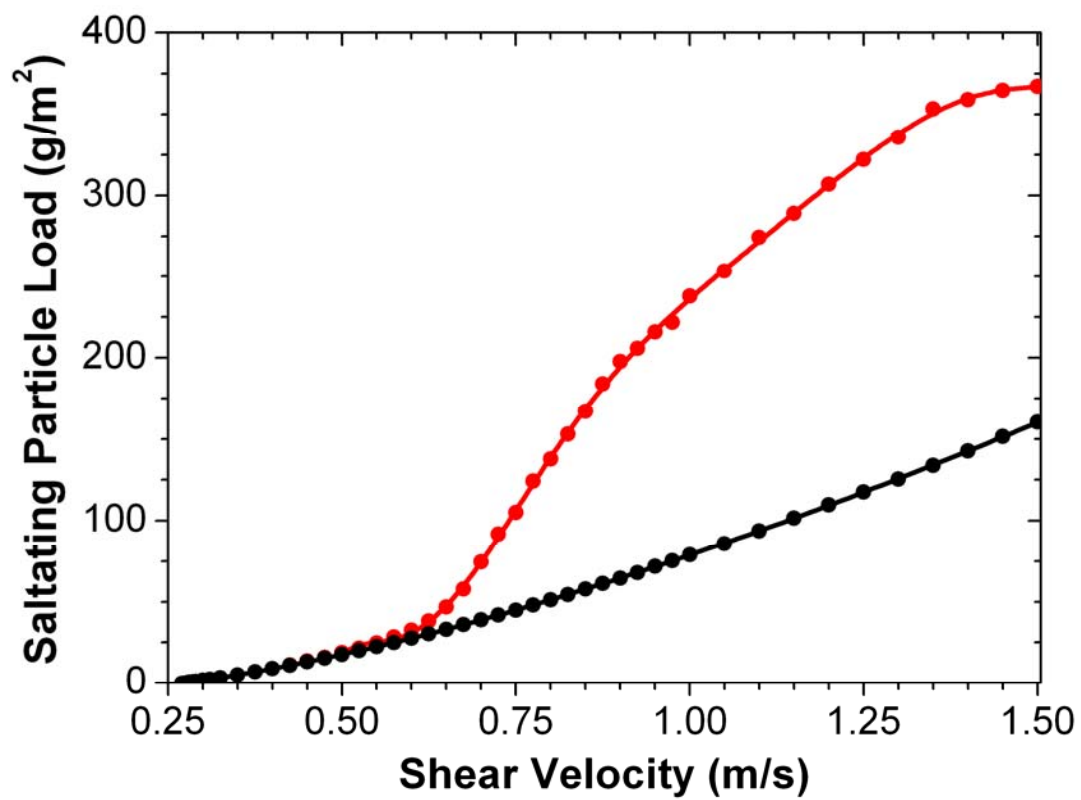

FIG. 4.

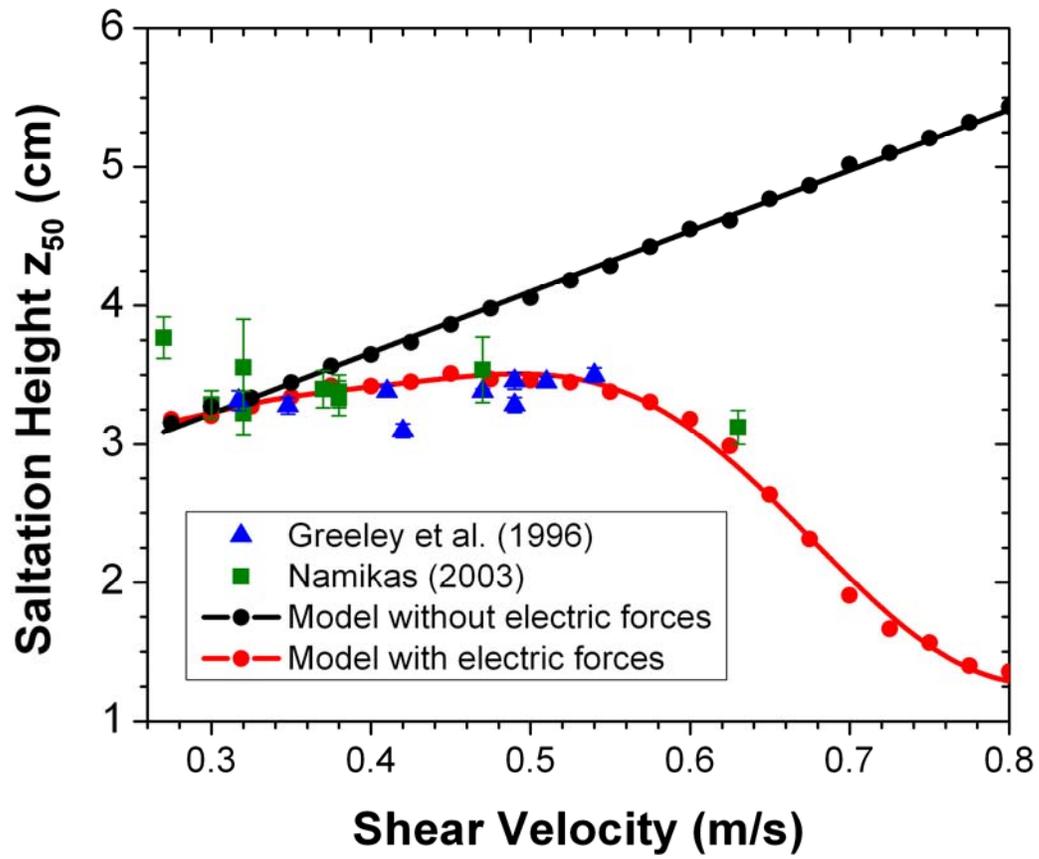

FIG. 5.

**FIGURE CAPTIONS**

FIG. 1. Schematic of saltation, showing the logarithmic wind profile $U(z)$ to the left of an idealized spherical sand particle propelled by the wind and bouncing along the surface. After lift-off from the surface, saltating particles obtain horizontal momentum from the wind, which is partially converted into vertical momentum after colliding with the surface and rebounding. The inset shows the force diagram of a negatively charged saltating particle over the positively charged soil surface.

FIG. 2. Comparison of measured (squares) and modeled (solid line) E-fields in saltation as a function of height. The measurements [6] were taken for winds of 4 – 12 m/s at 1.5 m height, which was estimated to correspond to an average shear velocity of 0.5 ± 0.1 m/s [2]. The soil particle size distribution was taken as typical for the broad top of a dune [20], where the measurements were made. The inset shows the dependence of the surface E-field on shear velocity for the size distribution reported in Ref. [13], with the dashed line corresponding to the electric lifting threshold [7].

FIG. 3. Vertical profiles of saltation mass flux measured in field experiments (squares [13] and triangles [14]), and compared to model results with and without electric forces (red and black solid lines, respectively). Both measured and modeled mass flux profiles are normalized by their total mass flux to simplify comparison. Results are shown for (a) low shear velocity (u* = 0.32 m/s), (b) medium shear

velocity ($u^* = 0.47$ m/s) and (c) high shear velocity ($u^* = 0.63$ m/s). Model results were obtained for the size distribution reported in Ref. [13].

FIG. 4. Mass load of saltating particles modeled with and without electric forces (red and black lines, respectively) as a function of shear velocity, for the size distribution reported in Ref. [13]. Electric forces cause the saltation particle load to increase at a given shear velocity by reducing the wind stress required to lift surface particles [7], and by reducing the average surface stress exerted by a single saltating particle [21].

FIG. 5. Dependence of the characteristic saltation height $z_{50}$ (see text) on the wind shear velocity. Classical saltation theory predicts that $z_{50}$ increases strongly with shear velocity [1,16], which our model also predicts when electric forces are not included (black circles). However, field measurements (squares [13] and triangles [14]) show that $z_{50}$ remains approximately constant. Inclusion of sand electrification in our model (red circles) resolves this discrepancy. The linear increase in $z_{50}$ without sand electrification is consistent with results from an independent numerical model [22]. Values of $z_{50}$ were obtained from Refs. [13,14] by fitting an exponential function to the measured vertical mass flux profiles, as described in Ref. [13]. Error bars represent the uncertainty in the fitting parameters. Model results were obtained for the size distribution reported in Ref. [13].

# Supplementary material to "Electrostatics in wind-blown sand"

**Jasper. F. Kok[1,2,*] and Nilton. O. Renno[1,2]**

[1]Applied Physics Program, University of Michigan, Ann Arbor, MI, 48109, USA.
[2]Atmospheric, Oceanic, and Space Sciences, University of Michigan, Ann Arbor, MI, 48109.

**Abstract**

In the main manuscript, we briefly describe our physically based saltation model and show that the inclusion of electric forces improves its predictions. Here, we give a more detailed description of the model and present its predictions of the horizontal and the height-integrated mass flux of saltating particles.

**Detailed model description**

A schematic overview of our model is given in Fig. S1. Following previous studies, we model saltation as the interplay of four processes [1]: (i) the motion of saltating particles, (ii) the modification of the wind profile through momentum transfer between the wind flow and saltating particles, (iii) the collisions of particles with the soil bed, and (iv) the lifting of surface grains by wind stress and by particles impacting the soil bed. In addition to these four critical processes, we also simulate the charging of saltating particles during collisions with other particles and the surface (see main manuscript). The effect of the resulting electric forces on particle motion and on the critical wind speed required to lift them from the surface [2] is calculated explicitly. Due to their interdependence, the wind profile, the electric field, and the saltating particles' motion, concentration, and charge, are calculated iteratively until steady state is reached (see Fig. S1).

*Equations of motion.-* As in previous studies [1,3,4], we model the motion of saltating particles in two dimensions. For simplicity, we neglect the effects of turbulence and mid-air collisions on particle trajectories, because these effects are relatively small for typical shear velocities [1,5].

In addition to including fluid drag and gravitational forces as in previous models [1,3-5], we also include the effects of electric forces (see main manuscript) on particle trajectories. Additionally, we include the fluid lift force, which is mainly important close to the surface [6]. The motion of saltating particles is then described by

$$ma_x = \frac{\pi D_p^2}{8} \rho_a C_d |V_R|(U - v_x)$$
$$ma_z = \frac{\pi D_p^2}{8} \rho_a \left[ -C_d |V_R| v_z + C_l \left( U_{top}^2 - U_{bot}^2 \right) \right] - mg + qE$$
(S1)

where $m$, $q$, and $D_p$ are the particle's mass, charge and diameter, respectively; $a_x$ and $a_z$ are the particle accelerations in the $x$ and $z$ directions; $v_x$ and $v_z$ are the particle speeds in the $x$ and $z$ directions; $V_R$ is the vector difference between the particle and wind velocities; $U$ is the horizontal wind speed, and $U_{top}$ and $U_{bot}$ are the wind speeds at the top and bottom of the grain, respectively; $\rho_a$ is the air density; $g$ is the gravitational constant; and $E$ is the electric field at the particle's position. We calculate the drag coefficient $C_d$ using the Reynolds number [7] of the saltating particle, assuming that it has the shape of natural sand particles, for which we take the characteristic length scale as $0.75D_p$ [6,8,9]. The lift coefficient $C_l$ is taken as $0.85C_d$ [10]. The equations of motion (S1) are numerically integrated for the calculation of the trajectories of the saltating particles.

*Wind profile.-* The initial trajectories of saltating particles are calculated using the logarithmic wind profile formed by a turbulent fluid flowing over a no-slip surface [11],

$$U(z) = \frac{u*}{\kappa} \ln\left(\frac{z}{z_0}\right), \tag{S2}$$

where $\kappa = 0.4$ is the von Karman constant, $z$ is the vertical distance from the surface, and $z_0 \approx 2D_p/30$ is the aerodynamic surface roughness [12]. As in previous investigations [1,3,4,6], the wind flow is assumed to be horizontal.

The initial wind profile given by (S2) is modified by the transfer of momentum and shear stress between the wind flow and saltating particles. The fluid shear stress is given by [5]

$$\tau_a(z) = \rho_a \left[\kappa z \frac{\partial U(z)}{\partial z}\right]^2. \tag{S3}$$

Outside the near-surface layer where saltation takes place (the 'saltation layer'), the fluid shear stress is a constant given by [5,13]

$$\tau = \rho_a u*^2, \tag{S4}$$

where $u*$ is the wind shear velocity. Owen [13] showed that the shear stress due to saltating particles ($\tau_p$) and the fluid shear stress in the saltation layer ($\tau_a$) sum to the fluid shear stress outside the saltation layer ($\tau$). That is,

$$\tau = \tau_a(z) + \tau_p(z), \tag{S5}$$

where [5]

$$\tau_p(z) = \sum_i m_i v_x^i(z) - \sum_j m_j v_x^j(z), \tag{S6}$$

and where the subscripts $i$ and $j$ respectively sum over all descending and ascending particles that pass the height $z$ per unit area and unit time.

Owen [13] further hypothesized that the fluid shear stress at the surface remains at 'a value just sufficient to ensure that the surface grains are in a mobile state'. That is,

$$\tau_a(0) \approx \tau_t, \tag{S7}$$

where $\tau_t$ denotes the threshold shear stress (critical value) necessary to initiate the motion of surface particles. Although Owen's hypothesis has not been fully verified [5], it has been widely used in analytical (e.g. [14,15]) and numerical (e.g., [4,16]) studies. We also use it, because it greatly simplifies the computation of both the wind profile and the concentration of saltating particles.

Since (S7) describes one of the main assumptions in our model, we briefly explain the reasoning behind it. Owen [13] argued that the fluid shear stress at the surface should remain at the threshold for particle entrainment because if the surface shear stress falls below the critical value for particle entrainment, fewer particles are entrained by wind. This reduces the transfer of wind shear stress and momentum between saltating particles and the wind flow, thereby increasing the surface shear stress back to its critical value. Conversely, if the surface shear stress exceeds the critical value, more particles are entrained, again restoring the surface shear stress to its critical value. Owen derived his hypothesis on the assumption that surface grains are predominantly lifted by aerodynamic forces. That is, by assuming that the ejection of surface grains by saltating particles impacting the soil bed ('splashing') is negligible. However, a numerical model developed by Anderson and Haff [1] predicts that the splashing of grains is the predominant method of particle entrainment in saltation. Nonetheless, they found that the fluid shear stress at the surface remains somewhat below the threshold shear stress. Raupach [14] argued that this result suggested that Owen's hypothesis might be more general.

Limited experimental support for Owen's hypothesis was given by Gillette et al. [17]. They found good agreement between field measurements and an analytical model of the aerodynamic roughness length in saltation that employs Owen's hypothesis [14]. The numerical study by Shao and Li [4] provided further support for Owen's hypothesis. This study found that splashing of grains only becomes important for large shear velocities, in disagreement with the earlier study of Anderson and Haff [1].

Using Owen's hypothesis, the concentration of saltating particles can be calculated in a straightforward manner. By combining (S5) and (S7), we find that the particle shear stress at the surface equals

$$\tau_p(0) = \tau - \tau_t. \tag{S8}$$

The concentration of saltating particles per unit area is then

$$N = \frac{\tau_p(0)}{\overline{\tau_{sp}(0)}} = \frac{\tau - \tau_t}{\overline{\tau_{sp}(0)}}, \tag{S9}$$

where $\overline{\tau_{sp}(0)}$ is the average shear stress that a single saltating particle exerts at the surface. With the concentration and trajectories of particles known, the particle shear stress at all heights ($\tau_p(z)$) is calculated. The fluid shear stress ($\tau_a(z)$) is then directly obtained through (S5), and used to numerically integrate (S3) to obtain the wind profile.

*Collisions with the soil bed.-* The collision of saltating particles with the soil bed is an essential process in saltation [1,5]. The experimental study of Mitha et al. [18] indicated that saltating particles have a probability of ~94 % of rebounding upon collision with the soil bed. The numerical study of Anderson and Haff [1] investigated the rebound probability ($P_{reb}$) in more detail and found that

$$P_{reb} = 0.95[1 - \exp(-\gamma v_I)], \tag{S10}$$

where $\gamma \approx 2$ m$^{-1}$s is a constant, and $v_I$ is the speed with which the particle impacts the surface. Equation (S10) is consistent with the result of Mitha et al. [18] for large impact speeds. Moreover, it predicts a decreasing chance of rebound for low impact speeds, as would be expected. We thus use (S10) to calculate the chance of an impacting particle to rebound from the surface.

Results of laboratory and numerical studies [1,3,18-22] are used to describe the velocity of particles rebounding from the surface as a normal distribution centered at 55 % (± 20 %) of the particle's impact speed, at an angle of 35º (± 15º) from horizontal.

The model calculates the charge transfer of rebounding particles with the soil surface using (3) and (4) in the main manuscript. The charge of particles that do not rebound from the surface is added to the total charge held by the soil surface.

*Lifting of surface particles.-* In steady-state saltation, the loss of saltating particles through (S10) must be balanced by the ejection of new particles from the surface, either through aerodynamic entrainment or through splashing. Using (S10), the model calculates the total number of particles that do not rebound, per unit area and unit time. In steady-state, this loss of saltating particles must be balanced by the sum of aerodynamically entrained and splashed particles. That is,

$$N_{\mathrm{nr}} = N_{\mathrm{a}} + N_{\mathrm{s}}, \tag{S11}$$

where $N_{\mathrm{nr}}$ is the number of non-rebounding particles per unit time and unit area, $N_{\mathrm{a}}$ (m$^{-2}$s$^{-1}$) is the number of aerodynamically entrained grains, and $N_{\mathrm{s}}$ (m$^{-2}$s$^{-1}$) is the number of splashed grains.

The aerodynamic entrainment of grains has been poorly studied, while the splashing of surface grains by impacting saltating particles has been investigated by several studies [1,3,22,23]. These studies found consistently that, for particles impacting a bed of similar particles, the number of ejected particles per collision ($n_{\mathrm{s}}$) varies approximately linearly with the speed of the impacting particle. Thus,

$$n_{\mathrm{s}} \approx A v_{\mathrm{I}}, \tag{S12}$$

where conservation of momentum allows the constant $A$ (m$^{-1}$s) to be rewritten as

$$A = a \frac{m_{\mathrm{I}}}{m_{\mathrm{bed}}}, \tag{S13}$$

where $a$ (m$^{-1}$s) is a constant, $m_{\mathrm{I}}$ is the mass of the impacting grain, and $m_{\mathrm{bed}}$ is the average mass of particles in the soil bed. For loose sand, we find that $a \approx 0.42$ m$^{-1}$s by fitting (S13) to the experimental results reported in Ref. [3]. Then,

$$N_{\mathrm{s}} = \sum_j n_{\mathrm{s}}^j = \frac{a}{m_{\mathrm{bed}}} \sum_j m_{\mathrm{I}}^j v_{\mathrm{I}}^j, \tag{S14}$$

where $j$ sums over all particles impacting the soil bed per unit area and unit time. Again using the result of laboratory and numerical studies, we describe the speed of splashed particles as a normal distribution centered at 8 % (± 4 %) of the speed of the impacting particle, at an angle of 55º (± 25º) from horizontal [1,3,23].

With $N_{\mathrm{s}}$ known, the number of aerodynamically entrained particles ($N_{\mathrm{a}}$) is found using (S11). A plausible assumption [1,4] is that the vertical component of the lift-off speed ($v_{\mathrm{a}}$) of these particles can be described by an exponential distribution

$$P(v_{\mathrm{a}}) = \frac{\exp(-v_{\mathrm{a}}/\overline{v_{\mathrm{a}}})}{\overline{v_{\mathrm{a}}}}, \tag{S15}$$

where we assume that the average vertical lift-off speed $\overline{v_{\mathrm{a}}} = \sqrt{8gD_{\mathrm{p}}}$. Neglecting non-gravitational forces, a particle with vertical speed $\overline{v_{\mathrm{a}}}$ would thus rise four grain diameters above the soil bed, which is a reasonable estimate [1]. We further assume that particles lift off at an angle similar to that of splashed particles (see above). The results reported in the main manuscript are mostly insensitive to the details of the parameterization of the lift-off of aerodynamically entrained grains.

The charge of surface grains lifted by aerodynamic entrainment or splashing is taken as their projected area ($\pi D_{\mathrm{p}}^2/4$) times the soil surface charge density ($\sigma$).

*Size distribution of saltating particles.-* Previous numerical models have focused mostly on saltation over soil beds made up of homogenous particles [1,3,4]. We here extend our model to include saltation over mixed soil beds, which is a better representation of natural conditions.

Several studies have found that the size distribution of saltating particles is similar to that of the parent soil [24-26]. A likely reason for this similarity is that saltation is critical to the formation of dune and beach sands, and these sands must thus contain particles that can be readily transported by saltation. More specifically, measurements by Namikas [9,27] show that the saltating particle size distribution is somewhat biased towards smaller particles, and shifts slightly towards larger particles with increasing shear velocity (Fig. S2). This probably happens because the wind shear velocity determines the size of the largest particles that can saltate. For simplicity, we assume that the saltating particle size distribution is identical to that of the parent soil for particles of 100-500 $\mu$m. This approximation is justified by the similarity between the size distributions of the saltating particles and the parent soil in this size range (Fig. S2 and Refs. [24-26]).

*Charge transfer during mid-air collisions.-* Although the effect of mid-air collisions on particle trajectories is neglected (see above), the model does account for charge transfer occurring during such collisions (see main manuscript). The chance $\Delta P_\lambda$ that a given saltating particle will collide with a particle in particle bin $j$ in the model time step $\Delta t$ is given by

$$\Delta P_\lambda(z,\vec{v}) = \Omega^j \rho^j \Delta v^j \Delta t , \quad (S16)$$

where $z$ and $\vec{v}$ are the position and velocity of the saltating particle, $\rho^j$ is the density (in m$^{-3}$) of particles in bin $j$ at the saltating particle's location, and $\Delta v^j$ is the average magnitude of the relative velocity with particles in bin $j$. The collisional cross section $\Omega^j$ is given by

$$\Omega^j = \frac{\pi}{4}(D_p + D_p^j)^2 , \quad (S17)$$

where $D_p^j$ is the diameter of particles in bin $j$.

The transfer of charge between the colliding particles is modeled using (3) and (4) in the main manuscript.

*Reduction of threshold shear velocity by electric forces.-* The laboratory experiments reported by Kok and Renno [2] show that electric forces provide an additional upwards force on surface grains, which aids the lifting of these particles by wind. They showed that the effect of electric forces on the threshold shear velocity is described by

$$u*_t = \sqrt{\frac{A_n}{\rho_a}\left(\rho_p g D_P + \frac{6\beta G}{\pi D_P} - \frac{8.22\varepsilon_0 E_{\text{surf}}^2}{c_s}\right)} , \quad (S18)$$

where $A_n \approx 0.0123$ is a dimensionless parameter that scales the aerodynamic forces, $\rho_p$ is the particle density, $\beta$ is an empirical constant that scales the interparticle force between sand grains and is on the order of $10^{-5} - 10^{-3}$ kg/s$^2$, $G$ is a geometric parameter that depends on the bed stacking and is of order 1, $\varepsilon_0$ is the electric permittivity of air, $E_{\text{surf}}$ is the surface electric field as calculated by (5), and $c_s$ is a scaling constant that accounts for the non-sphericity of soil particles. Our model uses (S18) to quantify the reduction of the threshold shear velocity by electric forces.

*Treatment of creeping particles.-* In wind-blown sand, the main mode of particle transport is saltation. However, a fraction of the mass transport occurs through particles striking the soil and pushing other particles in the direction of the wind flow. This near-bed transport of rolling and sliding particles is termed 'creep' [28]. Our model does not explicitly account for this mode of transport. However, most grains leave the surface (either through aerodynamic entrainment or splashing) with low speed, and will quickly settle back to the surface, as mathematically described by (S10). These grains can thus be considered to be transported through creep [18]. This interpretation is supported by the good agreement between predicted and measured vertical and horizontal mass flux profiles, especially close to the surface where creep is important (Figs. 3 and S3). Since electric forces reduce near-surface winds through (S18), transport by creep is generally enhanced by electric forces, because fewer ejected or aerodynamically entrained surface grains will acquire enough speed to prevent from quickly settling back to the surface.

**Additional model results**

In the main manuscript, we show that our model results are consistent with measurements of the vertical profiles of the saltation mass flux (Fig. 3). Here, we present additional tests of our physically based numerical model of saltation.

Modeled horizontal mass flux profiles show generally good agreement with field measurements (Fig. S3). The predicted height-integrated mass flux is also in good qualitative agreement with data from over a dozen wind tunnel studies and one field study (Fig. S4) [29]. In particular, the experimentally-determined peak in the dimensionless mass flux is well-reproduced by the model. To our knowledge, this peak has not been previously reproduced by numerical models and is often not reproduced by analytical models [29]. The predicted height-integrated mass flux appears larger than that measured by most experimental studies, which may be because sand collectors used in these studies have an efficiency of only ~50-70 % [30]. Moreover, the effects of interparticle collisions on particle motion are neglected in the present model. Sorensen and McEwan [31] hypothesized that interparticle collisions attenuate the saltation mass flux at high shear velocities. We intend to study the effect of interparticle collisions on saltation in a future publication.

The inclusion of electric forces brings the predicted height-integrated mass flux in better agreement with measurements (red line in Fig. S4). However, it is not clear whether electric forces are fully equilibrated in the finite length available in wind tunnels. For this reason, wind tunnel studies might be inadequate for studying fully-developed saltation. Experimental studies are required to resolve this issue.

**References**

[1]  R. S. Anderson, and P. K. Haff, Acta Mech. Suppl. **1**, 21 (1991).
[2]  J. F. Kok, and N. O. Renno, Geophys. Res. Lett. **33**, L19S10 (2006).
[3]  I. K. McEwan, and B. B. Willetts, Acta Mech. Suppl. **1**, 53 (1991).
[4]  Y. P. Shao, and A. Li, Bound.-Layer Meteorol. **91**, 199 (1999).
[5]  Y. P. Shao, *Physics and Modelling of Wind Erosion* (Kluwer Academic Publishers, Dordrecht, 2000).
[6]  R. S. Anderson, and B. Hallet, Geol. Soc. Am. Bull. **97**, 523 (1986).

[7] S. A. Morsi, and A. J. Alexander, J. Fluid Mech. **55**, 193 (1972).
[8] R. A. Bagnold, Geogr. J., **85**, 342 (1935).
[9] S. L. Namikas, Sedimentology **50**, 303 (2003).
[10] W. S. Chepil, Trans. Am. Geophys. Un. **39**, 397 (1958).
[11] L. Prandtl, in Aerodynamic Theory, edited by W. F. Durand (Springer, Berlin, 1935), Vol. III, p. 34.
[12] D. J. Sherman, Geomorphology **5**, 419 (1992).
[13] P. R. Owen, J. Fluid Mech. **20**, 225 (1964).
[14] M. R. Raupach, Acta Mech. Suppl. **1**, 83 (1991).
[15] Z. S. Li, J. R. Ni, and C. Mendoza, Geomorphology **60**, 359 (2004).
[16] G. Sauermann, K. Kroy, and H. J. Herrmann, Phys. Rev. E **64**, 031305 (2001).
[17] Gillette D. A., B. Marticorena, and G. Bergametti, J. Geophys. Res. **103**, 6203 (1998).
[18] S. Mitha, M. Q. Tran, B. T. Werner, and P. K. Haff, Acta Mech. **63**, 267 (1986).
[19] B. R. White, and J. C. Schulz, J. Fluid Mech. **81**, 497 (1977)
[20] P. Nalpanis, J. C. R. Hunt, and C. F. Barrett, J. Fluid Mech. **251**, 661 (1993).
[21] M. A. Rice, B.B. Willetts, and I. K. McEwan, Sedimentology **42**, 695 (1995).
[22] F. Rioual, A. Valance, and D. Bideau, Phys. Rev. E 62, 2450 (2000).
[23] M. A. Rice, B. B. Willetts, and I. K. McEwan, Sedimentology **43**, 21 (1996).
[24] D. A. Gillette, and T. R. Walker, Soil Science **123**, 97 (1977).
[25] D. A. Gillette, J. Adams, A. Endo, D. Smith, and R. Kihl, J. Geophys. Res. **85**, 5621 (1980).
[26] D. A. Gillette, and W. Chen, Earth Surf. Proc. Landforms **24**, 449 (1999).
[27] S. L. Namikas, J. Coastal Res. **22**, 1250 (2006).
[28] R. A. Bagnold, *The Physics of Blown Sand and Desert Dunes* (Methuen, London, 1941).
[29] J. D. Iversen, and K. R. Rasmussen, Sedimentology **46**, 723 (1999).
[30] R. Greeley, D. G. Blumberg, and S. H. Williams, Sedimentology **43**, 225 (1996); K. R. Rasmussen, and H. E. Mikkelsen, Sedimentology **45**, 789 (1998).
[31] M. Sorensen, and I. McEwan, Sedimentology **43**, 65 (1996).

**Figures**

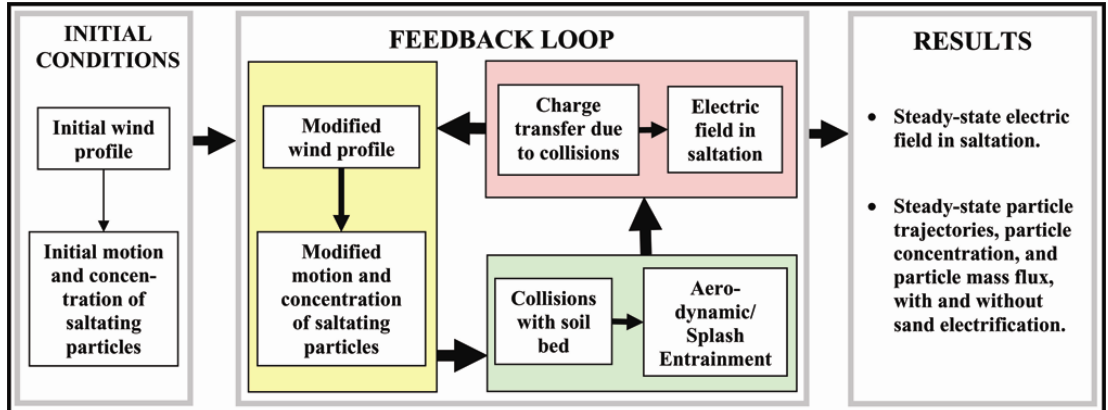

FIG. S1. Schematic diagram of our physically based saltation model. As in previous studies [1,3,4], saltation is modeled by explicitly simulating (i) particle trajectories, (ii) the absorption of momentum from the wind by these particles, (iii) the collision of particles with the soil bed, and (iv) the creation of new saltating particles by aerodynamic entrainment and the ejection of surface grains by saltating particles striking the soil bed. In addition to these four critical processes, our model also includes sand electrification and the effects of electric fields on saltation (see text and main manuscript). The steps indicated in the feedback loop are repeated until the changes in the saltation trajectories, the wind profile, and the electric field are smaller than a specified value in successive iterations.

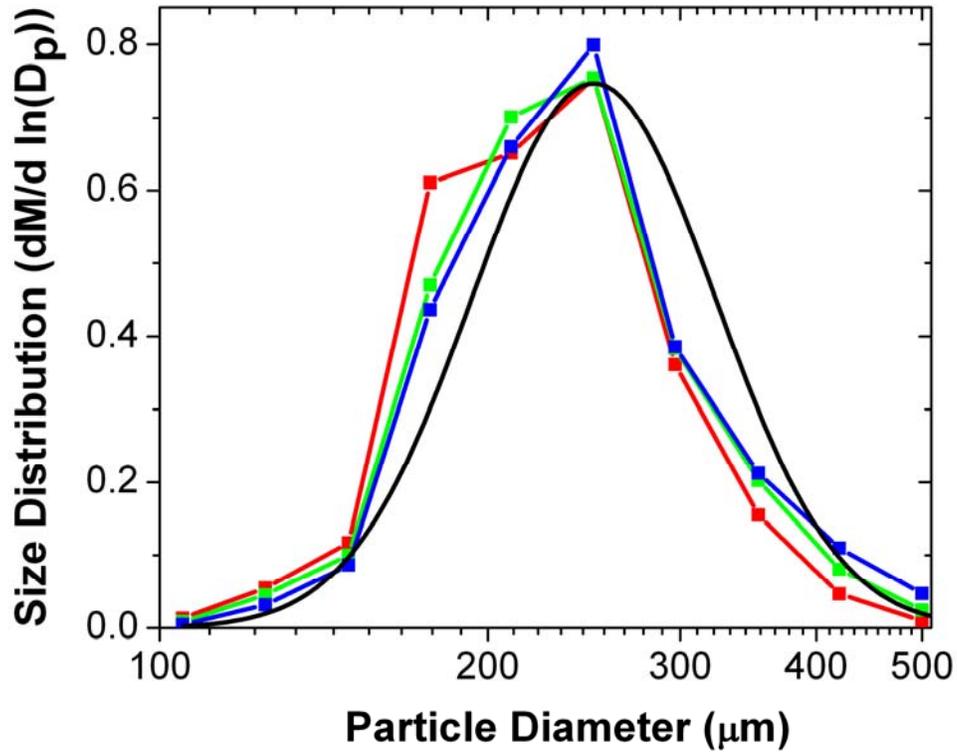

FIG. S2. Measured saltating particle size distributions for shear velocities of $u^* =$ 0.30, 0.36, and 0.55 m/s (red, green, and blue lines, respectively), where $M$ represents the fractional contribution to the size distribution. The parent soil size distribution (black line) is plotted for comparison [9]. Saltating particle size distributions were obtained from Fig. 3 in Ref. [27].

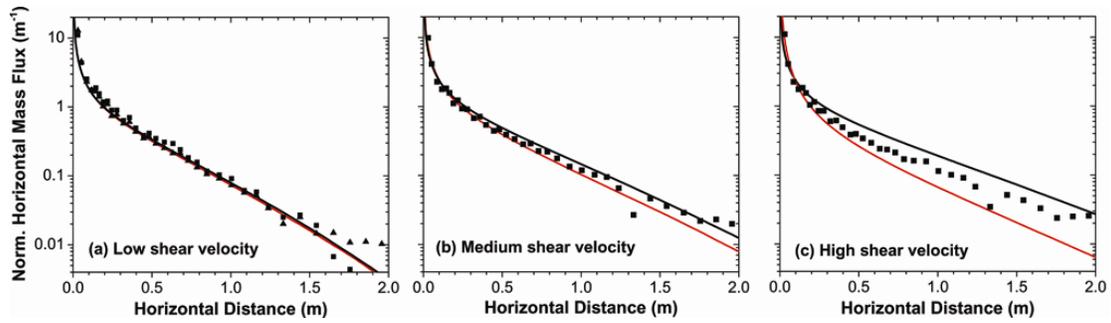

FIG. S3. Horizontal profiles of saltation mass flux measured in field experiments (squares [9]), and compared to model results with and without electric forces (red and black solid lines, respectively). Both measured and modeled mass flux profiles are normalized by their total mass flux to facilitate comparison. Results are shown for (a) low shear velocity ($u^* = 0.32$ m/s), (b) medium shear velocity ($u^* = 0.47$ m/s) and (c) high shear velocity ($u^* = 0.63$ m/s). Model results were obtained for the size distribution reported in Ref. [9].

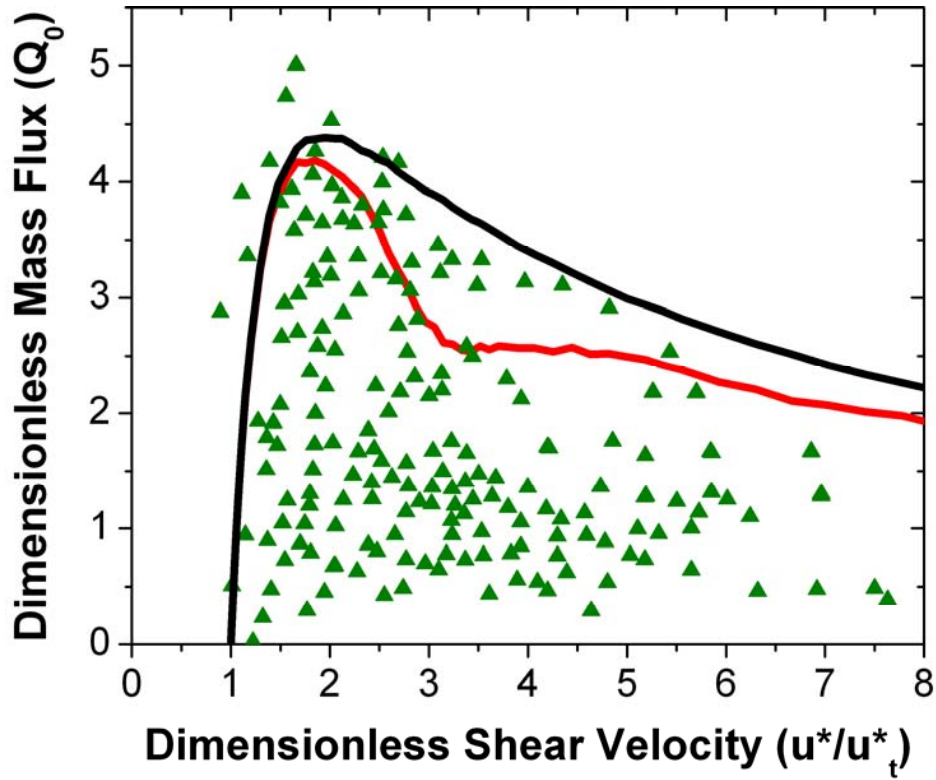

FIG. S4. Dimensionless mass flux $Q_0$ (see main manuscript) as a function of dimensionless shear velocity simulated with and without electric forces (red and black lines, respectively), and compared with results from more than a dozen wind tunnel studies and one field study (green triangles) compiled by Iversen and Rasmussen [29]. A peak in the experimentally determined dimensionless mass flux is apparent around $u^*/u^*_t \approx 2$, and is reproduced by the model. Model results were obtained for the size distribution reported in Ref. [9].